\documentstyle[11pt,newpasp,twoside]{article}
\def\cf{{\it cf.\ }}
\def\eg{{\it e.g.\ }}
\def\ie{{\it i.e.\ }}
\def\etal{{\it et al.\ }}

\def\edcomment#1{\iffalse\marginpar{\raggedright\sl#1\/}\else\relax\fi}
\reversemarginpar
\begin{document}
\title{The Phenomena of High Energy Astrophysics}
\author{R. D. Blandford}
\affil{130-33 Caltech, Pasadena, CA 91125 USA}
\begin{abstract}
A brief summary of some highlights in the study of high energy 
astrophysical sources over the past decade is presented. It is argued that the great progress
that has been made derives largely from the application of new technology to 
observation throughout all of the electromagnetic and other spectra and that, on this basis,
the next decade should be even more exciting.  However, it is imperative to observe
cosmic sources throughout these spectra in order to obtain a full understanding of their
properties. In addition, it is necessary to learn the universal laws that govern the macroscopic
and the microscopic behavior of cosmic plasma over a great range of physical 
conditions by combining observations of different classes of source. These two injunctions
are illustrated by discussions of cosmology, hot gas, supernova remnants
and explosions, neutron stars,
black holes and ultrarelativistic outflows. New interpreations of the acceleration 
of Galactic cosmic rays, the cooling of hot gas in rich clusters
and the nature of ultrarelativistic outflows are outlined. 
The new frontiers of VHE $\gamma$-ray astronomy,
low frequency radio astronomy, neutrino astronomy, UHE cosmic ray physics and gravitational 
wave astronomy are especially promising.
\end{abstract}
\section{Two Decades of High Energy Astrophysics}
There has been a wonderful decade of discovery in high energy 
astrophysics.
Accretion disks surrounding massive black holes in active galactic nuclei
(AGN)
have been traced into relativistic regimes using ASCA, Chandra and 
XMM-Newton (\eg Fabian 2002, Wilms \etal 2001).
Other disks around similar sources create ultrarelativistic
outflows, or jets, that have been directly imaged on scales from pc 
to Mpc using HALCA and the VLBA (\eg Junor, Biretta \& Livio 1999), 
and Chandra (\eg Wilson, Young \& Shopbell 2001) and, indirectly, probed on sub--pc scales
using EGRET (\eg Hartman \etal 1992)
and atmospheric Cerenkov telescopes 
(\eg Quinn \etal 1998). Similar jets and disks have been associated with 
Galactic X-ray binaries (XRB) and shown to exhibit Quasi-Periodic 
Oscillations (QPOs) using, especially, RXTE (\eg van der Klis 1998). 
Gamma Ray Bursts (GRB)  were 
placed at cosmological distances following observations by BATSE
(Meegan \etal 1992) which was confirmed by ground-based spectroscopy 
of X-ray afterglows discovered by Beppo-SAX (Costa \etal 1997, Metzger \etal 1997). 
These bursts also appear
to comprise collimated, ultrarelativistic outflows which eventually form
the afterglows discovered by Beppo-SAX and which, in turn, presumably 
evolve to form a small fraction of the supernova remnants (SNR) whose dynamics
and composition have been mapped at X-ray energies (\eg Canizares
2002).  The X-ray 
background was effectively associated with individual faint sources
by ROSAT (as confirmed by Chandra and XMM-Newton, \eg Hasinger 2002, Brandt
\etal 2002). Evidence of
the hot intergalactic medium (IGM)
has been found by FUSE (eg Tripp, Savage \& Jenkins 2000) and, more recently,reported with  
Chandra (\eg  Fang \etal 2002)
and we are now studying this same gas in rich clusters
of galaxies in fine spectroscopic detail (\eg Mushotzky 2002)   

These (electromagnetic) discoveries have been matched by great
discoveries in cosmic ray physics.  The atomic and 
isotopic composition of cosmic rays has been measured in exquisite detail by ACE
and the spectrum has been extended to ultra high energy (UHE)
by the AGASA and HiRes arrays. In addition, 
neutrino mass has been detected by painstaking work at Homestake, Kamionkande, SNO
and KamLAND (\eg  Eguchi \etal 2003).

These examples, which could surely be matched by a quite separate
list involving different sources and observatories, are, arguably, 
as far-reaching and of equal popular interest to the great discoveries
that have been made over a similar period in cosmology and extra-solar
planets. 

The coming decade should be no less exciting. Integral has just been 
launched.  Auger and Hess, which will
detect UHE cosmic rays at ZeV energy 
and VHE $\gamma$-rays at TeV energy with unprecedented sensitivity,
are just coming on line. Swift will study GRBs and produce a long overdue
hard X-ray survey. Astro-E2, has a planned 2005 launch 
and will perform high dispersion spectrosocopy of accretion disks etc. 
This will be followed quickly by GLAST which should be roughly
50 times a powerful as EGRET. There are ambitious plans to open up 
high energy neutrino astronomy by augmenting AMANDA and constructing 
IceCube. LIGO is already operational and it is hoped that it will
start gravitational radiation astronomy. There is also optimism that the 
space missions, Constellation-X/XEUS, LISA and EXIST will be 
started by the end of the decade.  

hat this impressive list of operating and 
planned missions also brings out is that high energy astrophysics is an 
integrating discipline. Sources are observable over $\sim70$
octaves of the electromagnetic spectrum (including the single octave
claimed by optical astronomers!) from $\la100$~MHz to $\ga10$~TeV.
If we look forward to gravitational and cosmic ray astronomy,
the spctrum expands to fill the interval from $\la100\mu$Hz gravitons
to $\ga1$~ZeV protons and the number of octaves doubles.
High energy sources are invariably nonthermal which implies that they
must be observed ``holistically''. Panchromatic campaigns to study 
AGN have been common for more than twenty years and, more recently, multi-wavelength 
observations have been the key to the study of GRB afterglows.

There is a second, integrating feature of high energy astrophysics
and this has been less appreciated 
and, so far, less exploited. This is that much of what 
we can observe depends upon a fairly small number of physical 
processes that we do not understand very well.
However, these processes should be source-independent.
Examples include the behavior of ultrarelativistic shock fronts,
the rates of thermalization and the thermal conductivity of 
hot magnetized plasma and the viscosity of shear flows.
Ultrarelativistic flows, in general, are seen in AGN jets, pulsar wind 
nebulae (PWN) and GRBs and there is every reason to undertake
comparative studies to understand their general, global, behavior.
The rate of particle acceleration and magnetic field generation
at a relativistic shock front ought to depend solely on the Mach 
number (or, equivalently, the Lorentz factor). Hot plasmas 
are observed in the laboratory, in the solar corona, in the 
inter-planetary, -stellar and -galactic media. Most of the important
transport processes should scale simply with density and in an
unknown, though universal manner with temperature. The effective
angular momentum, mass and energy transport in strongly, shearing
media, likewise probably depends on a set of elementary
principles. Here, numerical simulations are starting to be especially 
instructive. I expect that exploiting the wide variety of physical
conditions in cosmic sources to divine fundamental scaling relations
will be a major feature of high energy astrophysics research over the 
next decade. 
\section{The Cosmological Context}
The recent maturation of observational cosmology
has special implications for high energy astrophysics.  There 
now exists a cosmological framework in which to interpret 
the observations.  Although we do not understand much at all
about why this is the case, we do appear to inhabit
a universe with Hubble constant $\sim65{\rm km\ s}^{-1} {\rm Mpc}^{-1}$
age, $\sim14$~Gyr, a flat spatial geometry, and a current
composition of roughly 70~percent dark energy, 25 percent dark matter
and 5 per cent baryonic matter.  This knowledge, allows us to be 
much more quantitative when analyzing individual sources, especially 
when estimating pressures, densities, speeds etc.

High energy observations have contributed significantly to the 
development of this framework. The most important example is the 
measurement of the matter density by observing the X-ray emission
from clusters of galaxies. This has consistently given a value
of roughly $\Omega\sim0.3$ for a decade (predicated on the theory
of big bang nucleosynthesis) and it is now claimed that the measurement
error is better than ten percent (Allen, Schmidt \& Fabian 2002.  This same analysis 
led to an equally important value for the density
fluctuation normalization, $\sigma_8=0.7$. 

One example of the importance of being quantitative is in understanding
the growth of massive black holes in AGN.  The most recent 
determinations of the black hole mass density in contemporary galactic 
nuclei concludes that holes have to be assembled 
quite efficiently (with efficiencies $\epsilon\sim0.2$, Yu \& Tremaine 2002)
and therefore radiatively. Furthermore, the discovery of powerful 
quasars at redshift $z\sim6.5$ (Fan \etal 2001)
implies that the first black holes 
probably grew at a rate faster than the Salpeter rate,
simultaneous with the growth of the host galaxy.  This, in turn has
implications for the contribution of quasar ultraviolet radiation
(probably small) to the intergalactic, photoionizing radiation 
field. There is a good possibility that GRBs will be seen to even
greater redshifts than quasars and provide different probes of intervening material.

Another connection between high energy astrophysics and cosmology involves VHE
$\gamma$-rays which constrain the mid-far infrared background. A complementary 
constraint is provided by the highest energy cosmic rays whose range is 
similarly limited by photopion production on the microwave background. Here we know the
opacity very well and it it is the sources whose location is unknown, but must lie within
$\sim30$~Mpc at the highest energies detected. 
\section{Hot Gas}
The two best laboratories for studying 
transport processes at high temperature are rich galaxy clusters and supernova
remnants. Recent X-ray observations of both of these have been impressively
detailed and are still far from digested.  
\subsection{Clusters of Galaxies and the Intergalactic Medium}
Clusters are important 
cosmologically because as we have just seen,
they are thought to be large enough to provide
a fair sample of the baryons and because they are very convenient tracers
of the growth of large scale structure
that can be identified at large redshift through X-ray surveys, 
Sunyaev-Zel'dovich effect studies and weak lensing investigations. They are 
also important because they harbor the oldest galaxies at a given 
cosmic time and provide the best fossil record of the formation of the 
first galaxies. 

Clusters have become important physically because they can
teach us about the microphysical behavior of hot plasma. This has
been become central to attempts to resolve the ``cooling flow paradox''.
It has been known for a long while that the radiative cooling times 
of the gas at the centers of rich clusters is often shorter than the cluster
ages. It was then supposed that the gas would flow into the central
cD galaxies at rates as high as $\sim1000{\rm M}_\odot{\rm yr}^{-1}$.
What appears to be happening is that the gas starts to cool more 
or less as anticipated, but then it almost vanishes only to reappear at much lower
temperature radiating optical and ultraviolet emission lines copiously from gas with 
$T\sim2\times10^4$~K and density $n\sim100$~cm$^{-3}$ (\eg Peterson \etal 2003).

There are some important clues as to what is going on. 
The gas that is observed at high temperature
appears to be in thermal equilibrium and the isotherms are nested 
quasi-spherical surfaces. Another important clue 
has emerged from studies of clusters like the Perseus and Virgo clusters
that contain double radio sources. The X-ray emission from the areas of the
sky occupied by these radio sources is reduced, suggesting that the 
the strongly magnetized, relativistic plasma responsible for the radio 
emission does not, in practice, mix well with the cooler (\ie with temperature
$\sim10^8$~K!) plasma into which it
is expanding. This inference has been reinforced by the discovery of fossil 
radio sources, presumably associated with earlier phases of nuclear activity,
that are rising under buoyancy in the cluster gravitational field
(\eg Fabian \etal 2002). Some 
additional deductions have been made, rather more controversially. The gas 
immediately surrounding these bubbles is actually cooler than most of the
cluster gas. The existence of these large temperature gradients 
in rich clusters argues that the effective
mean free paths of hot electrons are smaller than given by
Coulomb scattering in unmagnetized plasma.
Most attempts to account for the thermal structure of clusters
have posited some form of heating to prevent the gas from cooling.
This seems rather unpromising. Piling gas up at a temperature where
it can radiate relatively efficiently does not seem a good way to make
it disappear! Also the radio sources, the most promising sources of 
distributed heating, do not appear to perform this function.

I would like to propose a rather different explanation for these observations.
The gas that accumulates in rich clusters has a very high entropy relative 
to the $\sim10^4$~K gas that was ionized when the universe was $\sim0.5$~Gyr old.
The most likely source of this entropy is passage through a strong shock 
front formed as galaxy-sized perturbations become nonlinear and drive gas with sound speed 
$\sim10$~km s$^{-1}$ together with speeds $\sim300$~km s$^{-1}$ (\eg Miniati \etal 2000). 
(Supernova explosions and expanding, double radio sources can also 
create strong shock fronts.) The  post-shock gas will have a 
density $\sim10^{-3}$~cm$^{-3}$, a temperature $\sim10^6$~K and a pressure  $\sim10^{-12}$~dyne
cm$^{-2}$.  Now a gravitationally-induced shock in the IGM should behave just like one
of similar Mach number ($M\sim30$) in the ISM (see below). This implies
that there should be a large, post-shock, cosmic ray partial pressure, roughly
$\sim0.3$ times the total pressure (\eg Miniati \etal 2000).
As the gas expands, following the 
passage of the shock, and as a consequence of the general
expansion of the universe, the cosmic ray pressure will become slightly more
important and may even dominate. However, if and when this gas collects into
a deep potential well formed by a rich cluster of galaxies, the gas will
be compressed and the 
cosmic ray pressure will decrease relative to that of the gas. The gas
that is observed to be cooling in centers of rich clusters
has a pressure similar to that in post shock gas and the $\sim$~GeV cosmic ray 
pressure is still likely to contribute about $\sim30$~percent of the total. 
These cosmic rays should make clusters into $\sim$~GeV $\gamma$-ray sources,
detectable by GLAST. Nonthermal emission from the electrons may have also been
seen in the extreme ultraviolet (\eg Durret \etal 2002). Finally, careful modeling
of relaxed clusters using X-ray, lensing and microwave background observations may 
lead to detection of a pressure deficit in the thermal gas.

When the cluster gas starts to cool, as it must eventually, it will compress by a 
factor of a few until the 
cosmic pressure dominates and resists further compression. The gas will then cool 
roughly isochorically and the inflow will be halted or at least seriously inhibited.
It is then possible for the cool gas to permeate the warm ($T\sim10^7$~K) gas 
and radiate away the internal 
energy contained in the warm gas. In principle, this can be very efficient. Suppose 
that the cool gas has a temperature $\sim10^5$~K, where its emissivity is maximized
(\cf Krolik 1999),
and a pressure of $\sim10^{-10}$~dyne cm$^{-2}$, typical of the center of a rich cluster.
$\sim3\times10^7$~M$_\odot$ of cool gas occupying a fraction $\sim10^{-5}$ of the 
volume suffices to radiate the missing soft X-ray power in the ultraviolet.

The problem is one of getting the energy from the hot gas to the cold gas fast enough.
The traditional approach is to suppose that the interface is a static, conductive atmosphere.
If the conductivity is dictated by Coulomb scattering then it scales $\propto T^{5/2}$
and the thermal contact is poor unless the cool gas is seriously 
overpressured with respect to the warm gas. However, this may not be the 
right description of the gas.
The situation is likely to be quite complex for
a variety of reasons. Firstly, the dynamical situation, may become 
unstable to Rayleigh-Taylor instablity. In addition, the galaxies that
move almost sonically through the cluster, will be followed by 
large turbulent wakes containing streaks of cool gas that has been stripped from 
galaxies.. 
Finally, the ongoing aggregation of large groups of galaxies will drive large
oscillations in the cluster gas. The gas is likely to end up quite well-stired
so that the warm gas flows past the 
cool gas on a timescale short compared with the conductive time so that the electrons may not 
be in local thermodynamic equilibrium.  To give a quantitative example, the electrons 
in the warm phase have speeds $\sim10,000$~km s$^{-1}$ and a Coulomb mean free path $\sim
30$~pc in the warm medium. It is therefore possible that the warm electrons 
may be brought closer to the cool gas by turbulent mixing than a mean free path 
and they are thereby able to
come into direct contact with it. With this arithmetic,the Coulomb
heating rate can now just balance the cooling rate of the cool gas.

A small quantity of cool gas, co-existing with the hot gas 
may then act as an effective heat sink, removing heat non-radiatively from the 
hot gas and radiating it away at a lower temperature.
Understanding the heat transfer is central to understanding the mass flow.  
In the simple theory of evaporation, an inward conduction of heat is balance by an outward
energy flux, $5Pv/2$. and the cloud evaporates. However what is envisaged here is that
that there is a volumetric heating which is roughly balanced by radiative cooling. If the
cooling exceeds the heating, there will be a steady condensation of hot gas onto the cool
cloud; if the heating dominates, there will be evaporation.  This may be self-regulating.
Clearly a much more careful investigation is called for to see if the above sketch has any
validity.
 
There has also been progress in studying the hot intergalactic
medium outside clusters
that recapitulates the progress that was made in understanding 
the interstellar medium following the launch of the Copernicus satellite.
FUSE has observed local IGM in emission at a temperature $\sim10^5$~K.
and both Chandra and XMM-Newton have reported detections of hotter gas
in both the local and the distant universe with temperature that may be as high
as $\sim5\times10^6$~K, although the interpretation of
these observations is not yet consistent. The state of the 
IGM is a good monitor of the development of both
large scale structure and stellar activity in the expanding universe.  This is 
because the IGM is probably only heated to a temperature $\sim1-2\times10^4$~K 
after reionization (which probably occurs when the universe is 
$\sim0.5$~Gyr old following the formation of the first massive stars,
Bromm, Coppi \& Larsen 1999).

\subsection{Supernova Remnants and the Interstellar Medium}

If we had never seen supernovae or their remnants
but had access to ultraviolet 
observations of hot stars, we would ask similar questions
of the interstellar medium. However, we now know that it is the 
explosions of massive stars and not gravitational action
that keeps most of the volume of the 
Galaxy at a temperature of nearly a million degrees. However, 
the details of how this happens are controversial. Most importantly, 
we do not understand how the bounding shock waves behave. We are not sure
how the post-shock electron and ion temperature depend upon the Mach 
number and how quickly these two components  equilibrate. The indications
are that the most of the energy flux is carried by the ions and that the 
electrons are heated at rates that reflect Coulomb scattering, but we need 
to be more quantitative.

The acceleration of Galactic cosmic rays has long been associated with 
supernova remnants. Most of the cosmic ray energy density is in the form
of $\sim1$~ GeV particles and models of the particle
acceleration suggest that typical high Mach number shocks transmit
a cosmic ray partial pressure that is $\sim0.1-0.5$ times the total
momentum flux. However, the observational evidence is confusing.
On the one hand, ASCA observations of SN1006 first showed the presence of a
nonthermal component in the X-ray spectrum which was taken 
as {\it prima facie} evidence that relativistic electrons were accelerated
by some shock waves. However, there are other remants where the 
TeV $\gamma$-rays might have been expected and these are not seen, suggesting
that the maximum energy to which the particles are accelerated is well 
below $\sim1$~TeV. 
Ion acceleration, can be detected though $\pi^0$ $\gamma$-rays.
which may be seen by Hess, VERITAS and GLAST and has been reported
by Enomoto \etal (2002), (but see Reimer \& Pohl 2002).

It is also of quite general importance to understand how magnetic
field behaves at shock fronts.  The nonthermal emission that is measured may
only require a simple compression of the pre-shock magnetic field so that its
energy density is ignorable. A much greater stretching of the field 
is likely to occur in the vicinity of the contact discontinuity where the 
ejecta from the stellar envelope interacts with the circumstellar 
medium.  A  combination of radio polarization, 
optical and X-ray observations 
should again furnish some quantitative answers.

As a further illustration of how important it is to mobilise all the high 
energy observations to try to figure out the principles 
which govern the behavior
of high temperature plasma, let me introduce a second ``cosmic ray'' 
paradox relating to the acceleration and propagation of Galactic cosmic
rays.  It is known that the observed cosmic ray spectrum has an
energy-dependence $N(E)\propto E^{-2.6}$ and that the energy dependence
of the observed spectra of secondary elements like Li, Be, B is steeper
than that of the primary particles, $N(E)\propto E^{-3}$.  
This implies that the grammage traversed by cosmic 
rays before they escape varies with energy as $\lambda\propto E^{-0.4}$
and, consequently, that the sourced spectrum of the primary particle
satisfies $S(E)\propto E^{-2.2}$, approximately. This is just what
is expected in simple views of shock accleration theory and was one of the 
strongest arguments for taking it seriously.  However, the observed
primary spectrum extends all the way up to the famous ``knee'' in the spectrum 
around 1~PeV. This is inconsistent with the observation mentioned above that
some supernova remnants 
only accelerate particles to much lower energy. It may also be incompatible
with the measured cosmic ray anisotropy. This can be estimated by 
the ratio of the column density of the local interstellar
medium $\sim1-2$~mg cm$^{-2}$ to the grammage $\lambda(E)$ traversed.  The 
observed anisotropy at high energy may also be too small to be compatible with the
standard power law model.

A possible resolution of this paradox is that individual
SNR accelerate cosmic rays with 
power law spectra up to energies in the interval $\sim0.1$~TeV
to $\sim1$~PeV so that their collective 
effect is to produce a quite convex source spectrum $S(E)$.  If this 
is now combined with a propagation model in which $\lambda(E)$
has a compensatory concave shape. In other words, the grammage does not
continue to decrease as a power law with increasing energy,
but saturates.  The quotient of the source
spectrum and the grammage then, coincidentally, leads to a single
power law spectrum. This leads to a prediction that the light element 
spectra should be concave. 
Whatever the true explanation, if we can understand on the basis of empirical
arguments like this how cosmic ray protons and electrons are accelerated 
by shock waves and how they propagate, then we can export this understanding 
to the IGM.  

\section{Supernovae}

The spectacular imaging spectroscopy of supernova remnants of 
many different types by Chandra is starting to address many of these questions
and should supply plenty of forensic evidence to enable us to reconstruct 
many of the details of the initial 
explosion, through studies of the distributions of the different
elements (Canizares 2002).   

SN 1987a was a touchstone for our understanding of core collapse supernovae
(\eg Michael \etal 2002).
It produced the first detected supernova neutrinos and a pleasing 
affirmation of the most fundamental principles of advanced stellar
evolution. However, it also provided some surprises.  The absence
of hydrogen from its intial spectrum led to its intial identification
with a Type Ia supernova. However it clearly is a variant on the Type II 
supernova known as Type Ib and is characterized by the absence of a hydrogen
envelope on the presupernova star. 

There were other surprises. The $\gamma$-rays 
were observed far earlier than expected suggesting that the ejecta become 
highly inhomogeneous very quickly. SN1987a is coming back into prominence 
as its blast wave is just now
encountering the equatorial ring of matter left behind by outflow
associated with the pre-supernova star.

Now GRB are currently most fashionably associated with Type Ic supernova in
which the helium envelope has also been stripped away. There are many 
puzzles raised by this model, foremost among them is the nature of the 
medium into which the relativistic blast wave, that forms the afterglow, 
expands. Consideration of what is observed in SN1987a could 
be quite instructive for models of GRBs. 
In particular we do not understand when neutron
stars are left behind and when black holes form. 
More fundamentally, we do not have a 
widely accepted, working model of how any core collapse supernova model 
works. It is possible that some of the features of GRB models could, 
contrariwise, provide the missing ingredient for the Type II case.

\section{Neutron Stars}

Most neutron stars are observed long after formation. They are seen 
as XRB and they appear to be over-represented in globular clusters and this 
is attributed to three body dynamical stellar exchanges and recycling.  
The relative formation rates of neutron stars and black holes
is also attracting some attention as three microlenses have been discovered
where the lens is argued to be an unseen star with mass well in excess
of the upper limit for a neutron star. This suggest that single black holes
might be quite common in the halo of our Galaxy. These holes will be rather
hard to detect by other means unless they have mass-losing companions ---
the accretion rate from the interstellar medium is probably 
too small to be of interest.

Neutron stars have long been regarded as cosmic laboratories where Nature 
allows us to witness experiments performed upon cold nuclear matter, that 
complement the experiments performed on hot nuclear matter at heavy ion 
colliders, exploring physical conditions that are otherwise inaccessible.
There have been several, recent reports, some of them quite controversial,
of new observations of extreme physical conditions associated with neutron 
stars.

Several soft gamma repeaters, exemplified by SGR 1806-20, have been identified
with magnetars, slowly rotating strongly magnetized neutrons stars.
In the case of SGR 1806-20, the surface field strength is measured as 
$\sim80$~GT while the rotation period is $\sim8$~s, ensuring that
the observed emission cannot be rotation-powered. This contrasts with the 
X-ray pulsar SAX J1808.4 3658 which has a measured period of $\sim2.5$~ms
approaching that of the fastest radio pulsar. 

Observations of the surface temperature are also important. In the case of 
RXJ1856.5 -3574, a temperature of $\sim0.7$~MK has been measured
and a very small radius $R\sim7$~km inferred, provoking the suggestion 
that this could be a quark star (Drake \etal 2002).  However, the case is quite unconvincing
as it depends upon an uncertain distance measurement and a particular spectral
decomposition (Braje \& Romani 2002).
For the central pulsar in the 821 year old supernova remnant 3C58,
the temperature is below 1 MK (Slane, Helfand \& Murray 2002). 
This is cooler than is computed
on the basis of standard, modified URCA cooling calulations. 
This could be an indication that there is a pion or a kaon 
condensate in the core of the star. Alternatively, the proton fraction
might be large enough to allow direct URCA cooling to take place
(\eg Yakovlev \etal 2002).

Finally, the X-ray source 1E1207.4 -5209 (Sanwal \etal 2002) exhibits two helium
absorption lines which may allow the neutron star surface potential 
to be measured.  As most neutron star masses have a pretty standard value
$\sim1.4-1.5$~M$_\odot$, this allows the radius to be guessed and the equation 
of state above nuclear density to be constrained.  
   
\section{Black Holes}

There has been equally exciting activity in the study of black holes.
Quasi-periodic oscillations (QPO) have been measured in several XRB, mostly 
associated with the accretion disk.  It is clear that there is a very rich
phenomenology to be understood, that encodes the mass and spin of the hole
and which should allow general relativity to be tested. Unfortunately,
there is still no widely accepted theory of QPOs that allows these 
identifications to be made. Normal modes of oscillation of specialized
disk models have been computed. However, there is not an easy way to see
how they can be excited and sustained at observable amplitudes and numerical 
simulations do not exhibit discrete modes like these. In addition,
it is highly unlikely that the oscillatory X-ray emission arises from 
the photosphere of the disk. Their spectra are far too hard 
for this. Instead, it seems more likely that they are produced in an active
corona. This requires there to be a strong coupling, presumably 
magnetic, to exist between the disk and the corona. If this is correct, then 
it certainly complicates the interpretation of QPOs.

There has also been a lot of attention  paid to the Fe K$\alpha$ lines, 
originally reported by ASCA in Seyfert galaxies. These are now seen
in XRB. Occasionally these lines are quite broad, which has been 
widely attributed to a combination of Doppler shift and gravitational redshift.
There are some puzzles.  Most lines are seen as narrow and it is not
known why and when they turn out broad.

The most extreme example is MCG 60-30-15, where the line can
extend from 2-6.5 keV.  This has been interpreted in terms of a model 
where spin angular momentum from the spinning black hole is extracted 
by the disks, with magnetic torques.(The energy released by steady, viscous 
disk accretion alone cannot account for the line profiles.) If this 
interpretation carries the day, then it supports the idea that relativistic
jets are powered by black hole spin.  

If additional evidence can be mustered
for massive black holes spinning rapidly, then this has some interesting
implications for their genesis.  The point is that if black holes are built up
by merging, as has often been proposed, then a retrograde capture of a small 
hole by a large hole will involve a larger transfer of angular momentum than
a prograde capture (\eg Hughes \& Blandford 2003). 
Therefore, black holes that grow by merging will
generally spin down rather than up and rapidly spinning holes are unlikely
to be assembled in this way. This is all consistent with the most recent 
comparisons of the density of local black holes needed to account for the 
$z\sim 2$ quasar light and the local hole mass density computed using the 
hole mass --- bulge velocity dispersion.correlation.  

\subsection{Adiabatic Accretion}
Much attention has recently been devoted to what happens when gas is supplied
to a hole at a rate much less than the Eddington rate. It has long
been known that if the viscosity is relatively large and the electron heating 
not much faster than Coulombic, then the radiative efficiency is low
and the gas will continue to heat up on an accretion timescale and will 
form a thick accretion disk or torus.  In fact, a similar outcome is possible
when  the gas supply rate greatly exceedcs the Eddington rate.  In this case,
the radiative efficiency will be high but the radiation will be trapped
by the accreting gas and the radiation pressure will support the thick disk.
Either case can be described as ``adiabatic accretion'', by analogy with 
the nomenclature used to describe supernova remnants.

There have been three models proposed to describe adiabatic accretion.  
Advection-Dominated Accretion Flows (Narayan \& Yi 1994)
are steady flows in which
all the mass that is supplied crosses the event horizon.
Convection-Dominated Accretion Flows (Quataert \& Gruzinov 2000), are non-stationary and the 
mass supply backs up. In ADiabatic Inflow--Outflow Solutions (Blandford
\& Begelman 1999), 
most of the energy that is released close to the 
hole is carried off in an outflow, usually, though not necessarily,
involving a mass losing wind. In this case, the mass accretion rate will be much 
less that than the mass supply rate. 

Perhaps the greatest challenge to these models is presented by our Galactic 
center (Baganoff \etal 2001).  Here we know the black hole 
mass ($\sim2.6\times10^6{\rm M}_\odot$) and can make an estimate of the mass accretion rate
($\sim10^{21}{\rm g\ s}^{=1}$). The bolometric luminosity is
$\sim10^{36}{\rm erg \ s}^{-1}$ and so the radiative efficiency relative
to the mass supply is only $\sim10^{-6}c^2$. Most of the 
power emerges in the sub mm part of the spectrum. The X-ray emission
appears to be rapidly variable with large flares developing on timescales
less than an hour. The X-ray spectrum is 
steep and presumably nonthermal implying an upper 
limit on the density of gas close to the hole.  The 
large linear polarization at wavelengths
$\la1$~mm are similarly indicative that the density is low.  These
observations are strongly suggestive that the rate of mass accretion 
onto the hole is much less than the rate of supply, implying that most of the 
supplied mass is driven off in a wind, powered by the small fraction that
accretes onto the hole.This is entirely natural as the torque that 
transports angular momentum outward in an accreting flow also transports
energy so as to unbind an adiabatic flow. Provided that there is a either
a means of creating entropy at the disk surface, as happens in the solar wind,
or large scale magnetic fields are present (as is also true 
of the solar wind), then  outflows are to be expected.

Similar outflows are to be expected in the high mass accretion rate 
case and these are presumably responsible for the observed broad 
absorption lines  that are observed from many quasars.  These principles
should also apply to accretion onto Galactic black holes, for example
GRS 1915+115 (\eg Mirabel \& Rodriguez 2002). 
\section{Ultrarelativistic Outflows}
The most dramatic phenomena that are observed in high energy astrophysics 
are associated with the highest energy particles and
the most nonthermal spectra.  These, in turn, have been 
associated with ultrarelativistic outflows. I would now like to be provocative and 
suggest that we may have been seriously misinterpreting
most of these flows at least in recent years and that much older interpretations
may have been much closer to the truth (\cf Blandford 2002). 

\subsection{Pulsar Wind Nebulae}
Let me start with pulsar wind nebulae, like the 
Crab Nebula.  These are powered by central, spinning, magnetized neutron stars
and there is no dispute that the mechanical spin energy of the star is 
steadily converted into an electromagnetic Poynting flux that carries
energy into the magnetosphere. To order of magnitude we can associate
a flux $\Phi\sim10^{14}$~Wb with the open field lines of a typical pulsar
and if the angular frequency is $\Omega\sim100$~rad s$^{-1}$, the 
induced EMF is $V\sim\Omega\Phi\sim10^{16}$~V. We 
can think of this driving a current flow through the magnetosphere.
Under electromagnetic conditions, the ``load'' in the circuit will be 
$Z\sim100$~$\Omega$, and therefore the current will be
$I\sim V/Z\sim10^{14}$~A. The power dissipated in the load ---
essentially the pulsar luminosity --- is $L\sim VI\sim1^{30}$~W. 

Where is this load located? The
conventional view is that this electromagnetic energy flux is somehow converted
into a particle energy flux, perhaps in the vicinity of the light cylinder and
probably comprising electron-positron pairs. This is the location of the 
load. The Lorentz factor of the wind
speed has been estimated to be as high as $\sim10^6$. This fluid outflow
is then supposed to pass through a strong shock where its momentum flux 
matches the ambient nebular pressure and where relativistic particles
are re-accelerated.and, perhaps, magnetic field is regenerated.

However, what is seen in the recent X-ray observations 
of pulsar wind nebulae is very surprising. Polar jets are quite common 
(\eg Helfand 2001) and the 
ring-like structures that are observed appear to be confined
to the equatorial plane. (It is tempting to associate the
moving, ring-like features, especially in the Vela supernova remnant,
with large glitch activity as occurs roughly once per two years.)  
This tells us that accretion disks
are not necessary to create a jet morphology. It also tells us that 
if there really are fluid outflows that energize the ray emission, then 
these are concentrated at the poles and in the equatorial plane. 

What I think these observations
are, instead, telling us is that the current does not dissipate
near the light cylinder but flows out into the nebula. There are relatively strong arguments
that suggest that large amplitude waves with periods equal to the 
rotation period will become nonlinear and unstable and therefore it
is simplest to assume that the currents well beyond the light 
cylinder are primarily conduction currents.  However, this is not required.

We then have a picture of a pulsar wind nebula as a giant electrical
circuit with current flowing out (in) along the poles and in (out)
in the equator. The current completes at the slowly expanding surface of the 
nebula where there is a contact discontinuity against the shocked interstellar
medium.  Associated with these currents is a magnetic field that is '
largely toroidal that can be thought of as spun off by the central star.
The energy flow in the nebula is given by the Poynting flux, $\vec E\times\vec B$.
The electric field distribution is basically poloidal and derives from space charge
distributed along with the currents. If the pressure and the inertia of the plasma 
in the nebula can be ignored, and this is the appropriate approximation to make 
under electromagnetic conditions,then the electromagnetic field will be force-free
\ie  $\rho\vec E +\vec j\times\vec B=0$.  The electromagnetic setup
can be much more complex than this simple model.  There will probably 
be currents and space charge flowing throughout the nebula
and the pressure and inertia of the 
plasma in the nebula may well be important, but the simple model is sufficient 
to fix ideas.can be ignored. 

Such a configuration is generically unstable (just like fluid 
jets).  Typically, pinches, kinks and, especially, helices develop
around line currents in plasmas. Likewise, sheet currents, like those in the
equatorial plane, are subject
to tearing mode instability. Usually, this is regarded as a fatal defect for a model.
However, I would argue that it is an attractive feature of the present proposal. 
This is because it is possible that the 
nonlinear development of these instablities is responsible for the 
electrical resistance in the circuit and for the X-ray emission 
that is observed along the poles and in the equatorial plane.
(Note that the source of the power that 
is dissipated in this manner is the magnetic energy 
stored in the nebula; it does not flow along the jet, but as Poynting 
flux from the 
body of the nebula to where the current flows.)
One possible way that this can happen is that the macroscopic
instability drives a wave turbulence cascade that ultimately 
is dissipated at some inner scale through the acceleration 
of relativistic electrons.

There are a variety of predictions associated with this model. The two most
direct are that the if the electrons are accelerated close to the 
currents, then there should be spectral evidenced for aging as the particles 
diffuse away from the putative acceleration sites.Secondly the linear polarization
ought to reflect the underlying magnetic field geometry. 

\subsection{AGN Jets}

The X-ray images of extragalactic (and also Galactic) 
jets are no less striking.and present 
a similar choice.  It now seems to be generally accepted
that jet power derives from electro-/hydromagnetic stress applied 
on the black hole spacetime and the gas that orbits it. The details 
are just as contentious as with pulsars. In round numbers,
a powerful radio source, like Cygnus A, will generate an EMF
$V\sim300$~EV, a current $I\sim3$~EA and a power $L\sim10^{39}$~W.
Again, it is commonly presumed that
the circuit closes to the black hole and a fluid jet is collimated and 
launched. (Observations of M87 suggest that 
the collimation happens within $\sim100m$, Junor, Biretta \& Livio 1999.) 
The various features that are seen using VLBI in the compact, relativistic jets are 
usually identified
with internal shocks --- the nonlinear development of velocity gradients 
associated with either the source or instabilities (\eg Blandford 
\& K\"onigl 1979). Their measured,
outward, superluminal motion is then that of a shock front and the Doppler
beaming is that of the downstream flow, which moves more slowly 
than the shock front.

The jets themselves are impressively well-collimated. 
Even more remarkable is the discovery that they are X-ray bright 
along their length.  In the case of sources like M87, it is argued that 
X-rays are due to synchrotron radiation. This imples that the particle acceleration
must be occuring all the way along the jet's length because the cooling times 
of the $\ga10$~TeV electrons are very short. 
This, in turn, implies that the acceleration cannot occur
at internal shocks because strong shocks must be separated 
by much more than the synchrotron cooling lengths. 

By contrast, if we adopt the electromagnetic model, the 
``jets'' delineate the current which flows all the way 
along the jet to the hot spots and then back to the central black hole mostly
along the periphery of the source. The extended radio lobes contain
a reservoir of magnetic energy that can supply the emitting regions with 
energy in addition to the Poynting flux of energy flowing along the jet.
(It is helpful to think of magnetic energy as moving with a speed 
equal to $\vec E\times\vec B/B^2$ which can be 
arbitrarily close to $c$.  Disturbances can therefore be observed
moving with apparent superluminal speed just like the disturbances in 
fluid jets.) 
Continuous X-ray emission along the jet causes no problem 
because the current is continuous and ohmic
dissipation / particle acceleration can occur all the way along it.

\subsection{Gamma Ray Bursters}

Finally, consider, Gamma Ray Bursters (GRBs)
(\eg M\'esz\'aros 2002).  These come in two basic types
with short and long duration.  Only the latter class has been well-studied.
Several ``long'' bursts have been associated with afterglows that have
been observed from radio frequencies to X-ray energies. Many of the models 
of gamma ray bursts basically involve some form of electromagnetic induction.
A field of $\sim100$~GT associated with a rapidly spinning
stellar mass black hole or neutron star can induce an EMF $V\sim30$~ZV
and a power $\sim10^{43}$~W. The source is typically active 
for $\sim100$~s so the total energy of the burst is $\sim10^{45}$~J.
Observations of achromatic breaks in the afterglow light curves have been 
used to argue that the explosion is not isotropic but instead beamed within
a solid angle, typically $\sim0.1$~sterad. In other words, GRBs are 
relativistic jets too.

The conventional view of GRBs is, once again, that the Poynting flux is 
quickly and continuously transformed into a radiation-dominated fluid with a 
high entropy per baryon, typically $\sim10^6k$. This ``fireball'' 
(Cavallo \& Rees 1078) is 
collimated into a pair of anti-parallel jets, perhaps within a collapsing
massive star. As the flow accelerates along these jets, the internal 
energy contained in the radiation and pairs is transformed into the 
kinetic energy of the protons and by the time the photons can escape,
the ions are moving with Lorentz factors $\Gamma\sim300$. Small 
velocity gradients, induced at the source, steepen into internal shocks
at a radius $\sim10^{11}$~m and $\gamma$-rays are emitted 
as synchrotron radiation.  Most of the afterglow emission is formed 
at radii $\sim10^{15-16}$~m at the external shock 
that precedes the spreading, decelerating jet. 

However, if the energy is released electromagnetically,
it is quite hard to understand how entropy can be created so quickly.
The potential differences are so large that the vacuum is, in effect, a 
perfect conductor so that the invariant $\vec E\cdot\vec B=0$.
If the other invariant $c^2B^2-E^2$ is negative, then it will be possible
to transform into a frame where there is a pure electric field which will
instantly discharge; if it is positive, there will exist frames 
with a pure magnetostatic field, where nothing will happen. The problem
is essentially the same as that of the pulsar wind and I argue that it is quite
reasonable that the escaping power be predominately electromagnetic.

There is a second, serious concern with the fireball model. High 
Lorentz factors are necessary in the jet in order to avoid pair 
production by the escaping $\gamma$-rays. However, these high 
Lorentz factors are equivalent to high Mach numbers 
($M\ga300$) and it is very hard 
to see how these can be formed naturally and be sustained while, at the 
same time, dissipating much of their kinetic energy through internal 
shocks.

The electromagnetic model of GRBs posits that the energy remain in 
low entropy,
electromagnetic form all the way out to the $\gamma$-ray 
emission region which can be located around $\sim10^{14}$~m. The flow is
no more than mildly supersonic, depending upon the plasma loading. 
By the time this radius is reached, the electromagnetic field 
will be confined to a thin, relativistically expanding shell, pushing
a blast wave out into the surrounding medium. Instabiliites in this shell
will ultimately be responsible for the particle acceleration and the 
GRB.

Whatever one's view of the relative merits of fluid and electromagnetic
models of ultrarelativistic outflows, and perhaps the truth lies between the
two extremes described above, it is clear that there is a convergence in 
the study of pulsar wind nebulae, AGN/XRB jets and GRBs.       

\section{Physics at the Frontier}

High energy astrophysics is a young and relatively immature field.
It owns much 
of the remaining unexplored ``discovery space'' in contemporary astronomy.
Two examples of this discovery space are the extremes of observation
of the electromagnetic spectrum. At the high end,
there are already about ten 
TeV sources, while at the low end of $\la50$~MHz radio astronomy there are 
essentially no sources.
Neutrino astronomy claims only two cosmic
sources so far, the sun and SN1987a.  
Even the venerable field of cosmic ray physics may be on the threshold of 
becoming cosmic ray astronomy. Finally, as many of the most interesting 
high energy sources
are ultimately black holes and neutron stars, the exciting 
field of gravitational wave astronomy --- perhaps a decade away from birth
--- is inextricably linked to high energy astrophysics.  These are the 
frontiers and I consider them in turn.

Atmospheric Cerenkov techniques are being used to detect $\gamma$-rays in the
GeV--TeV range. These are important as both sources and as probes. 
Persistent sources have been identified with pulsars, blazars and 
supernova remnants and in each case are likely to provide the best approach
we have to understanding the fundamental nature of these sources. The big 
issue in pulsars is to locate the source of the emission. Is it close
to the stellar surface or at an outer gap, much closer to the light cylinder?
As the pulses are phase-resolved, we can also relate the $\gamma$-ray emission 
site to that of the radio, optical and the X-ray emission which is also 
not yet certain.  Blazars are identified with ultrarelativistic jets emanating
from massive black holes in the nuclei of elliptical galaxies. Here 
the big question is to understand whether the jets comprise
ultrarelativistic protons, that interact with either the radiation 
field or the background plasma, or if they are electron-positron 
pairs/electromagnetic.
TeV sources should be far more plentiful in the latter case.
The combination of GLAST and telesopes like Hess and VERITAS ought to 
be able to sort this out, as they should also be able to sort out the 
details of cosmic ray acceleration in supernova remnants, as discussed
above.   

At the other end of the spectrum lies the domain of low frequency radio astronomy
(Kassim \& Weiler 1990). 
The ionosphere precludes regular observing much below $\sim30$~MHz and it is 
necessary to fly large antennae in space to explore this part of the spectrum.
We can be confident that the sources exist.  They are necessarily nonthermal. 
The ultimate limit in frequency is the plasma frequency of the solar wind,
around $\sim30$~kHz, giving us another ten octaves to explore. There will of course
be limitations. Interstellar propagation will lead to irreversible smearing of 
any pulsar pulses, for example. 

The great advantage of neutrino astronomy is that allows one to see into the 
densest regions, even through nuclear density.  The recent successes 
in the SNO and Kamionkande and KamLAND experiments have verified that the three
types of neutrino have mass and can mix into each other. This vindicates
the standard solar model although the low energy neutrino spectrum remains 
to be measured. 

The next step in neutrino astronomy is to detect sources at much higher energy.
This is the province of projects like AMANDA, IceCube and ANTARES.  
The prime candidate sources of ultra high energy neutrinos are 
blazars and GRBs. In many respects, these searches are complementary
to the UHE $\gamma$-ray searches. Success in the former will suggest
that ultrarelativistic outflows comprise mainly protons.
Failure, and there is no guarantee that there will be {\it any} detectable 
UHE neutrino sources,
will favor electromagnetic/pair models UHE.
neutrino astronomy has the advantage that we can see the universe up 
to $\sim$~EeV energies. By contrast, the universe becomes opaque 
to $\gamma$-rays above  $\sim$ TeV energies through absorption 
by the infrared background.  

The cosmic ray frontier is undoubtedly at the very highest energy. The
situation is now quite confused. The EeV cosmic ray spectrum can be
measured through detecting the atmospheric nitrogen fluorescence that 
they create and by recording the muon showers on the ground. Unfortunately 
different spectra are being reported --- probably a consequence of 
calibration errors.  What is at stake is the source redshift. If the
cosmic rays derive from cosmological distances, then they should cut off
above $\sim100$~EeV due to photopion production by the cosmic microwave 
background.  If this ``GZK'' cutoff is not seen then their sources
probably lie within $\sim30$~Mpc and their overall mean luminosity 
density approaches that of $\sim$ GeV cosmic rays.  Even more
tantalizing are studies of the angular distribution of these particles. 
There are a few close groupings of particles --- five doubles and 
one triple --- on the sky. The statistical significance is low,
but if these are substantiated as permanent sources, then almost
all proposed models of UHE cosmic ray origin will be ruled out, with the 
conspicuous exception of the front runner, radio sources associated with 
dormant massive black holes in AGN. Both of these questions should be answered
by Auger.

Finally the last and most challenging frontier is that of gravitational
radiation.  Studies of binary pulsars have confirmed that the weak field 
calculations of wave emission are correct to an accuracy $\sim0.002$.
This is the most impressive confirmation of the general 
theory of relativity to date. This allows us to 
compute waveforms etc from strong field sources, according
to the rules of general relativity with confidence, although not
facility. However, it does not rule out the possibility that additional fields
are attached to singularities, for example, or that the relationship between 
curvature and stress-energy is more subtle.  Furthermore, it is at least
logically possible that in a universe where this relationship
{\it appears} to fail on the comsological scale, 
there could be some side effects affecting the propagation of metric
perturbations.  In short there are pretty good reasons in physics
to test the theory of gravitational radiation although the most likely 
outcome will be to vindicate, once again, the genius of Albert Einstein.

Gravitational radiation astronomy is largely unkown territory. There are
however assured sources - Galactic binary dwarfs and extragalactic
neutron star coalescences.  Of more interest, 
though, are coalescing black holes in cosmologically distant galactic nuclei,
though here the source rates are very difficult to estimate with confidence
and measuring them would tell us much of interest about galaxy evolution.
There are two classes of detectors. Ground-based facilities, like LIGO,
TAMA and VIRGO, will seek stellar sources like supernova and 
compact object mergers. mHZ, space-based facilities, such as
LISA, will target massive black hole signals from behind a binary white 
dwarf-generated foreground. The technical challenge of achieving
the sensitivities necessary to measure waves from assured
sources should not be understated. It may well take more than another
decade to reach them. However,    
there have been few regions of the electromagnetic spectrum where the sources
have turned out to be as we imagined. So, it would 
be remarkable if gravitational wave astronomy, or any of the other 
four frontiers turned out to be as I have just described. 

As I hope that this brief introductory essay makes clear, the past decade has 
been a remarkable one in the observation of 
high energy astrophysics. The present one promises even more. Let me conclude by 
re-emphasizing the two important principles
with which I began this article. Firstly, in order to exploit high energy observations to the
full it is necessary to adopt a source-based, rather than observatory-based approach to 
the study of cosmic sources. Almost by definition, high energy sources are nonthermal and emit 
throughout the electromagnetic and other spectra and spectrally chauvinist interpretations of 
their behavior are incomplete. Secondly, much of the physics of these sources is unknowable 
working from first principles. While we should be confident in our application of fundamental
principles, such as the conservation laws of mass momentum and energy and the 
whole edifice of general relativity, and in our understanding
of elementary processes such as those described by atomic and nuclear astrophysics
and quantum electrodynamics, much of what we see depends upon the collective behavior of plasmas
and the mysteries of MHD. As such, it is encumbent upon us to develop theories 
of these subjects by studying all sources from the solar corona to GRBs so as to derive
empirical laws which we can then try to relate to numerical simulation. The physics should be 
common and knowable.

I am confident that subsequent contributions to this meeting will report great advances
that can be interpreted in terms of both of these principles.
\section*{Acknowledgements}
I thank the organisers of this conference for their extremely gracious hospitality and
many participants for constructive criticism of some of the ideas discussed above.
I am also indebted to my collaborators, especially Mitch Begelman,
Max Lyutikov, Andrew MacFadyen and Peng Oh. 
Support under NASA grant 5-7007 is gratefully acknowledged.

\end{document}